\documentclass[aps,twocolumn,showpacs,prd,eqsecnum,10pt]{revtex4-1}
\usepackage{amsmath,amssymb,color,colordvi}
\usepackage[pdftex]{graphicx}
\usepackage{dcolumn}
\usepackage{bm}
\usepackage{dsfont}
\usepackage{color}
\usepackage[english]{babel}

\begin{document}

\title{Quantum mechanical response to a driven Caldeira-Leggett bath}

\author{Hermann Grabert$^{1}$ and  Michael Thorwart$^{2,3}$ }

\affiliation{$^1$Physikalisches Institut, Universit\"at
Freiburg, Hermann-Herder-Stra{\ss}e 3, 79104 Freiburg, Germany \\
$^2$I.\ Institut f\"ur Theoretische Physik, Universit\"at Hamburg,
Jungiusstra{\ss}e 9, 20355 Hamburg, Germany \\
$^3$The Hamburg Center for Ultrafast Imaging, Luruper Chaussee 149,
22761 Hamburg, Germany}

\begin{abstract}
We determine the frequency-dependent response characteristics of a quantum system to a driven Caldeira-Leggett bath. The bath degrees of freedom are explicitly driven by an external time-dependent force, in addition to the direct time-dependent forcing of the system itself. After general considerations of driven Caldeira-Leggett baths, we consider the Rubin model of a chain of quantum particles coupled by linear springs as an important model of a quantum dissipative system. We show that in the presence of time-dependent driving of the chain, this model can be mapped to a quantum system which couples to a driven Caldeira-Leggett bath. The effect of the bath driving is  captured by a time-dependent force on the central system, which is, in principle, non-Markovian in nature.  We study two specific examples, the exactly solvable case of a harmonic potential, and a quantum two-state system for which we assume a weak system-bath coupling. We  evaluate the dynamical response  to a periodic driving of the system and the bath. The dynamic susceptibility is shown to be altered qualitatively by the bath drive: The dispersive part is enhanced at low frequencies and acquires a maximum at zero frequency. The absorptive part develops a shoulder-like behavior in this frequency regime. These features seem to be generic  for quantum systems in a driven Caldeira-Leggett bath. 
\end{abstract}

\date{submitted to Phys.\ Rev.\ E, March 26, 2018}


\maketitle

\section{Introduction}\label{sec:one}

It is a common approach to open quantum systems \cite{Weiss2008,Bogolyubov45,Magalinskii59,Vernon63,Caldeira1981,Caldeira1983,Caldeira1983b,Leggett1987,Grabert1988} to identify few dynamical degrees of freedom, which are of interest to an observer, as the 'central system' with a few modes of well-defined frequencies, while the quantum statistical fluctuations are characterized by a continuous frequency distribution and are subsumed into the notion of an infinitely large 'environment' with many degrees of freedom each of which is not of explicit interest. The environment is mostly assumed to be in its thermal equilibrium state forming then a thermal bath. To probe the dynamical modes of the central system, an external time-dependent field is usually applied, and its frequency-dependent response yields information about transitions between the system states and the relaxation and dephasing properties due to the interaction with the bath. For strong external time-dependent fields, the characteristic features of the central system can be qualitatively modified \cite{GrifoniDrivenTunneling} and the forcing can even be used to control the time evolution of the system of interest \cite{BrummerQuantumControl}. Quite generally, the interplay of the quantum statistical fluctuations and a continuously acting driving force leads to the formation of a steady state of the driven dissipative quantum system of interest which is in general non-thermal. 

In order to model the response of the quantum statistical system to external forces, it is often sufficient to limit the coupling of the driving field to the central system alone. This is, in fact, an assumption often tacitly made when the response of dissipative quantum systems is studied \cite{Weiss2008,Caldeira1983b,Leggett1987,Grabert1988,GertIngold}. For instance, for atoms in optical lattices, single atoms can be addressed by laser fields \cite{Blatt} and the manipulation of single spins, e.g. by voltage pulses, has been demonstrated \cite{Hanson}. In appropriate cases, the assumption that the driving force couples solely to the central system is certainly reasonable. 
However, in particular for nanoscale condensed matter systems, it is often unavoidable as a matter of principle that the external driving field also couples to the environmental degrees of freedom. For instance, in superconducting nanocircuits, a superconducting coplanar wave guide is lithographically fabricated in close proximity to the superconducting loop forming a qubit, with a mutual capacitive coupling \cite{Devoret,Makhlin,Blais}. These subsystems cannot be addressed externally completely independently.  Another example is the orientation or manipulation of polar molecules of a solvent by THz fields \cite{Zalden,Santra} which affect the dielectric environment of a central molecular dipole and may lead to strong effects on its molecular polarizability. It is the main purpose of this work to provide a general theoretical framework to study the impact of a continuous distribution of explicitly driven environmental bath modes which couple to a central quantum system of interest.  

Based on a microscopic system-bath model, we extend the standard theory of damped quantum systems to driven systems. The external time-dependent driving force is assumed to couple to both the quantum system under consideration and the continuous distribution of environmental bath modes. To be explicit, we generalize the Caldeira-Leggett model \cite{Caldeira1981,Caldeira1983,Caldeira1983b} of a linear bath described by a set of harmonic oscillators bilinearly coupled to the central system. In particular, we add an explicitly time-dependent driving term to the bath Hamiltonian which couples to the displacements of the individual bath oscillators (dipolar bath driving). The driving term describes  an external forcing interacting with a system variable and the coordinates of the bath modes. The central quantum system thus experiences two types of driving, the direct driving and the indirect driving mediated by the driven continuous bath. To elucidate the consequences of a driven quantum bath, we extend in this work two paradigmatic models of quantum dissipation, the damped quantum particle moving in a potential field and the dissipative two-state system. The extended models are shown to be fully characterized by two memory functions: first, the well-known friction kernel which is associated with the effective spectral density of undriven bath modes, and second, the bath driving induces an effective force with a delay kernel which  incorporates also the coupling strengths of the environmental modes to the applied driving force. Specific results are given for a driven Rubin model and a modification thereof with a central two-state system. 

The Rubin model \cite{Rubin1,Rubin2,Weiss2008,GertIngold} of quantum dissipation roots in the theory of a one-dimensional harmonic lattice of point-like masses which are connected by linear springs. All but the central mass are equal and are assumed to be much lighter than the heavy-mass central particle. A diagonalization in terms of normal modes is possible and the velocity autocorrelation function of the central heavy mass can be obtained. Rubin noticed that, upon the assumption of a canonical distribution of the coordinates and the momenta of the light masses at initial time, the autocorrelation coincides with that of a Brownian particle coupled to a harmonic thermal bath. The spectral density of the bath can be calculated explicitly in terms of the microscopic parameters of the harmonic lattice and differs from the often used form of an Ohmic bath which is closely connected to Markovian behavior  \cite{Weiss2008}. In particular, the damping kernel of the Rubin model decays in an oscillatory manner of the form $\sim J_1(\omega_D t) / t$, where $J_1(x)$ is the Bessel function of the first kind and $\omega_D$ is the highest frequency of the bath oscillators or, resp., the frequency cut-off. Only at low frequencies, $\omega \ll \omega_D$, the behavior is Ohmic. Hence, the Rubin model yields a physical example for an open quantum system whose dissipative dynamics is non-Markovian in general, with a memory time being of the order of $1/\omega_D$. Thus, the interplay of the external bath driving and the intrinsic oscillations of the bath correlations can be expected to yield interesting effects.   

We find that the response of the Rubin model is significantly modified by the driven bath. In general, the dispersive and the absorptive parts of the susceptibility are shifted in frequency compared to the case with an undriven bath. Moreover, both parts acquire pronounced spectral weight in the low-frequency sectors. The dispersive part develops a maximum at zero frequency, while the absorptive part displays a significant shoulder at low frequencies.  We find this modified dynamical response, both, for the case of a central particle in a harmonic potential and for a two-state system. In general, the effect of the bath driving is entirely accounted for by a time-dependent force on the central system whose momentary value depends on the prehistory of the system-plus-bath and, thus, is, in principle, non-Markovian. Interestingly enough, the time-local time-dependent bath drive can be used to induce non-Markovian effects and to control the frequency-dependent response of the central system to the drive. 
\section{Quantum Brownian motion in a driven quantum bath}\label{sec:two}
We consider a quantum particle moving in a potential field and coupled to a bath of harmonic oscillators. The familiar Caldeira-Leggett model \cite{Caldeira1981,Caldeira1983,Caldeira1983b} is supplemented by a coupling term to an external time-dependent force interacting with the quantum particle directly and, in particular, also with the continuum of the bath oscillators.

\subsection{Model Hamiltonian}
The microscopic Hamiltonian of the entire driven system may be written as
\begin{equation}\label{HBM}
H_{\rm tot}(t) = H_S + H_R + H_{\rm int} + H_{\rm ext}(t)\, ,
\end{equation}
where
\begin{equation}\label{HSy}
H_S = \frac{p^2}{2M} + V(q)
\end{equation}
is the Hamiltonian of a particle of mass $M$ moving in a potential $V(q)$.
\begin{equation}\label{HR}
H_R = \sum_n \left( \frac{p_n^2}{2m_n} + \frac{1}{2}m_n\omega_n^2 x_n^2\right)
\end{equation}
describes a bath of harmonic oscillators, and
\begin{equation}\label{Hint}
H_{\rm int} = -q\sum_n c_n x_n + q^2\sum_n \frac{c_n^2}{2m_n\omega_n^2}
\end{equation}
is the interaction between system and bath. Here, the last term is a counter term assuring that the potential $V(q)$ is not altered by the coupling. These three parts of the Hamiltonian define the standard Caldeira-Leggett model \cite{Caldeira1981,Caldeira1983,Caldeira1983b} of a particle in a potential coupled to an environment with the effective spectral density
\begin{equation}\label{Jeff}
J(\omega)=\frac{\pi}{2}\sum_n \frac{c_n^2}{m_n\omega_n}\delta(\omega-\omega_n) \, .
\end{equation}
This quantity incorporates the spectral density of states of the bath modes and their coupling strengths to the quantum particle. 
The driving term
\begin{equation}\label{Hext}
H_{\rm ext}(t) = - \left( d_0 q  + \sum_n d_n x_n\right) F(t)
\end{equation}
adds the coupling to an external driving force $F(t)$. Here, we have allowed for a direct coupling of the external field to the quantum particle with coupling constant $d_0$. An additional feature of the extended model is a coupling of the driving field $F(t)$ to the bath oscillators with coupling constants $d_n$. Driven bath modes have only been addressed very recently \cite{Grabert_2016,Reichert2016} and the model studied here will be shown to provide a microscopic basis for the effects of driven quantum baths presented there.

\subsection{Evolution equations}

The Heisenberg equations of motion for observables $A$
\begin{equation}
\dot A(t) = \frac{i}{\hbar}\left[H_{\rm tot}(t),A(t)\right]
\end{equation}
take the form 
\begin{eqnarray}\label{Heomx}
\dot x_n(t) &=& \frac{p_n(t)}{m_n} \, ,
\\\label{Heomp}
\dot p_n(t) &=& -m_n\omega_n^2 x_n(t) + c_n q(t) + d_n F(t)
\end{eqnarray}
for the bath degrees of freedom. These equations are linear and can easily be solved for factorizing initial conditions \cite{Weiss2008} to yield
\begin{eqnarray}
x_n(t) &=& x_{n0}\cos\left(\omega_n t\right) + \frac{p_{n0}}{m_n\omega_n}\sin\left(\omega_n t\right)
\\
&& \nonumber +\frac{1}{m_n\omega_n} \int_0^t ds \sin\left[ \omega_n(t-s)\right] 
\left[c_n q(s) + d_n F(s)\right] \, ,
\end{eqnarray}
where $x_{n0}$ and $p_{n0}$ are the Heisenberg operators at the initial time $t=0$.

When these evolution equations are combined with the Heisenberg equations of motion of the variables of the quantum particle
\begin{eqnarray}
\dot q(t) &=& \frac{p(t)}{M} \, ,
\\
\dot p(t) &=& -\frac{\partial V(q(t))}{\partial q(t)} \\ \nonumber
&&\quad +\sum_n\left( c_n x_n(t) -\frac{c_n^2}{m_n\omega_n^2}q(t)\right) + d_0 F(t)\, ,
\end{eqnarray}
one finds after a partial integration
\begin{eqnarray}\label{le}
&&M\ddot q(t) + M\int_0^t ds \gamma(t-s) \dot q(s) +\frac{\partial V(q(t))}{\partial q(t)}\\
&&\nonumber \quad = \xi(t) + d_0 F(t) +\int_0^t ds\, \lambda(t-s)F(s) \, .
\end{eqnarray}
Here, 
\begin{equation}\label{gamma}
\gamma(t) = \frac{1}{M}\sum_n \frac{c_n^2}{m_n\omega_n^2}  \cos\left( \omega_n t\right) 
\end{equation}
is the damping kernel describing time-retarded friction of the particle. It can be written in the usual way in terms of the effective spectral density of bath modes (\ref{Jeff}) as
\begin{equation}\label{gammaJ}
\gamma(t)=\frac{2}{M}\int_0^{\infty}\frac{d\omega}{\pi}\frac{J(\omega)}{\omega}\cos(\omega t)\, .
\end{equation}
The random force operator
\begin{eqnarray}
\xi(t) &=&\sum_n c_n \left[\left(x_{n0}-\frac{c_n q_0}{m_n\omega_n^2}\right)\cos\left(\omega_n t\right) \right.\\
\nonumber &+& \left. \frac{p_{n0}}{m_n\omega_n}\sin\left(\omega_n t\right)\right]
\end{eqnarray}
depends on the initial values $x_{n0}=x_n(0)$, $p_{n0}=p_n(0)$ of the bath operators. We also use the notation $q_0=q(0)$.

The coupling to the force $F(t)$ in Eq.\ (\ref{Hext}) leads to the additional terms on the right-hand side of Eq.~(\ref{le}) accounting for the direct influence of the external force and the effective driving mediated by the bath. This latter effect is also time-retarded and characterized by the force delay kernel
\begin{equation}\label{lambda}
\lambda(t)=\sum_n \frac{c_n d_n}{m_n\omega_n} \sin\left( \omega_n t\right) \, ,
\end{equation}
which incorporates the spectral density of bath modes and their coupling strengths to the quantum particle and to the external force, respectively.

\subsection{General effect of driven bath modes}

The analysis presented so far allows for some general conclusions. The Caldeira-Leggett model of linear quantum dissipation is valid whenever the thermal bath is only weakly perturbed by the system dynamics so that the effect of the interaction between system and bath on the bath can be described within linear response theory. The linear response of a thermal bath can always be modeled in terms of a reservoir of harmonic oscillators with proper effective spectral density. The restriction to linear response of the bath does not mean that the influence of the bath on the system is weak, i.e., weak damping, since the combined effect of the large number of bath modes on the system can be very strong even if each of the bath modes is perturbed only weakly. 

When now an external force is perturbing the bath so that the bath is not driven beyond the range of validity of linear response theory for the bath, the system will be affected by the driven bath through a time-retarded impact of the external force which is fully described by an effective external force
\begin{equation}\label{Feff}
F_{\rm eff}(t) = d_0 F(t) +\int_0^t ds \, \lambda(t-s)F(s)\, ,
\end{equation}
as can be seen from the evolution equation (\ref{le}).
Here, the first term comes from the direct coupling of the system to the external force while the second term is due to the driving of the bath modes which affects the system dynamics by a term depending both on the coupling constants $c_n$ between system coordinate and  bath oscillators and on the coupling constants $d_n$ between the bath modes and the external force as exemplified in Eq.~(\ref{lambda}). 

The existence of an effective driving force $F_{\rm eff}(t)$ for driven Caldeira-Leggett baths irrespective of the specific dynamics of the central system is the main outcome of this work.  With this insight, the whole formalism of linear quantum dissipation \cite{Vernon63,Grabert1988,Weiss2008} can now be extended to externally driven baths. This includes the path integral formulation and the quantum master equation approach for damped particles in potential fields. We will next discuss a specific case of quantum Brownian motion.


\section{Driven Rubin Model}\label{sec:three}
\begin{figure}
\includegraphics[width=0.4\textwidth]{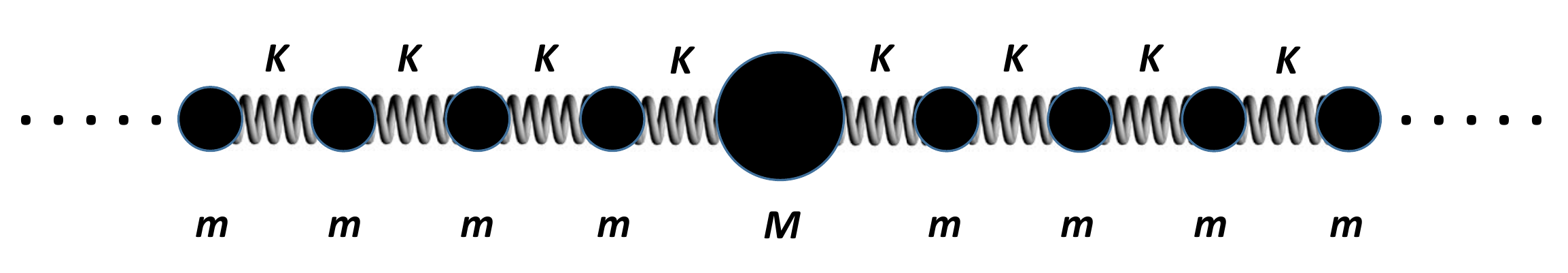}
\caption{\label{fig1} The Rubin model. A heavier central particle of mass $M$ moving in a potential field (not shown) is attached to two semi-infinite chains of lighter particles of mass $m$ which are connected by linear springs with spring constant $K$.}
\end{figure}

A well-known model  of quantum (and classical) Brownian motion has been introduced by Rubin \cite{Rubin1,Rubin2}. It describes a heavy  particle of mass $M$ coupled to two semi-infinite harmonic chains formed by particles of mass $m$ coupled by harmonic springs with spring constant $K$, see Fig.~\ref{fig1}.
 The corresponding Hamiltonian is given by
\begin{eqnarray}\label{Hru}
H_{\rm Rub}&=&\frac{p^2}{2M}+V(q)  +\sum_{S=L,R}\Bigg\{\frac{K}{2}\left(q-q_{S, 1}\right)^2  \\ \nonumber
&&+\sum_{k=1}^\infty\left[ \frac{p_{S, k}^2}{2m}+\frac{K}{2}\left(q_{S, k+1}-q_{S, k}\right)^2\right]\Bigg\} \, .
\end{eqnarray}

\subsection{Relation with Caldeira-Leggett model}
The Rubin model is not explicitly of the Caldeira-Leggett form. Using the transformation
\begin{equation}\label{trans}
q_{S, k}= \sqrt{\frac{2}{\pi}}\int_0^\pi d\psi \,\sin(k\psi) x_S (\psi) \, ,
\end{equation}
one obtains from the Hamiltonian (\ref{Hru}) 
\begin{eqnarray}\label{Hru2}
&&H_{\rm Rub}=\frac{p^2}{2M}+V(q)  +\sum_{S=L,R}\Bigg\{\int_0^\pi d\psi\, \Bigg[ \frac{p_S(\psi)^2}{2m}  \\  \nonumber
&&\ 
+2K \sin^2\!\left(\frac{\psi}{2}\right)\!x_S(\psi)^2-\sqrt{\frac{2}{\pi}}K \sin(\psi) q x_S(\psi)\Bigg]+\frac{K}{2}q^2 \Bigg\} \, .
\end{eqnarray}
Here, $p_S(\psi)$ is the momentum conjugate to $x_S(\psi)$ and we have made use of the identity
\begin{equation}
\sum_{k=-\infty}^\infty e^{ik \psi}= 2\pi \sum_{k=-\infty}^\infty \delta(\psi -2\pi k) \, .
\end{equation}
Mapping the left and right semi-infinite chains to negative and positive values of $\psi$, the Hamiltonian takes the form
\begin{eqnarray}\label{Hru3}
&&H_{\rm Rub}=\frac{p^2}{2M}+V(q)  +\int_{-\pi}^\pi d\psi\, \Bigg[ \frac{p(\psi)^2}{2m}  \\  \nonumber
&&\ 
+2K \sin^2\!\left(\frac{\psi}{2}\right)\!x(\psi)^2-\sqrt{\frac{2}{\pi}}K\vert \sin(\psi)\vert q\, x(\psi)\Bigg]+Kq^2 \, , 
\end{eqnarray}
which is a Caldeira-Leggett model with continuous summation index $\psi$. The frequencies of the bath modes are given by
\begin{equation}\label{om}
\omega(\psi)=2\sqrt{\frac{K}{m}}\, \left\vert \sin\left(\frac{\psi}{2}\right)\right\vert
=\omega_D \left\vert \sin\left(\frac{\psi}{2}\right)\right\vert \, ,
\end{equation}
where we have introduced the Debye frequency of the chains
\begin{equation}
\omega_D=2\sqrt{\frac{K}{m}} \, .
\end{equation}
It describes the maximum available frequency in the bath. The coupling constants are
\begin{equation}\label{cpsi}
c(\psi)=\sqrt{\frac{2}{\pi}}K\vert \sin(\psi)\vert \, .
\end{equation}
Note that the counter term
\begin{equation}
q^2\int_{-\pi}^\pi \! d\psi\,\frac{c(\psi)^2}{2m\,\omega(\psi)^2}
=q^2\int_{-\pi}^\pi \! d\psi\,\frac{K\sin^2(\psi)}{4\pi \sin^2(\psi/2)}=Kq^2
\end{equation}
corresponds indeed to the last term in Eq.~(\ref{Hru3}). Hence, the Rubin model is a particular case of a Caldeira-Leggett model with a spectral density of the form
\begin{eqnarray}\label{JR}
&&J(\omega)=\frac{\pi}{2}\int_{-\pi}^\pi d\psi\, \frac{c(\psi)^2}{m\,\omega(\psi)}\delta(\omega-\omega(\psi)) \\ \nonumber
&&\ =
\frac{K^2}{m\omega}\int_{-\pi}^\pi d\psi\, \sin^2(\psi)\delta(\omega-\omega(\psi))=
m\omega\sqrt{\omega_D^2-\omega^2} \, ,
\end{eqnarray}
confined to the interval $0\leq \omega \leq \omega_D$. This leads to the damping kernel
\begin{eqnarray}
\gamma(t)&=&\frac{2m}{\pi M}\int_0^{\omega_D} d\omega\, \sqrt{\omega_D^2-\omega^2}\,\cos(\omega t) 
\\ \nonumber
&=&\frac{ m\omega_D}{M}\frac{J_1(\omega_D t)}{t} \, ,
\end{eqnarray}
where $J_1(z)$ is a Bessel function of the first kind. The slow oscillatory algebraic decay of the damping kernel is a key feature of the Rubin model and illustrates the non-Markovian character of the temporal correlations of the bath-induced fluctuations. 
For later purposes we also note that the Laplace transform of $\gamma(t)$ is given by
\begin{equation}
\hat\gamma(z)=\frac{2m}{\pi M}\int_0^1 \! dx\frac{z\sqrt{1-x^2}}{(z/\omega_D)^2+x^2} 
=\frac{m}{M}\left(\sqrt{z^2+\omega_D^2} \!-\! z\right)\, ,
\end{equation}
where we have made use of the relation
\begin{equation}
\int_0^1 dx\frac{\sqrt{1-x^2}}{a+x^2}=\frac{\pi}{2}\left(\sqrt{\frac{1+a}{a}}-1\right) \, .
\end{equation}

\subsection{Driving force}

We next extend the model and add a driving force $F(t)$ which couples to all particles of the model Hamiltonian (\ref{Hru}). We choose a coupling term of the from
\begin{equation}\label{Hextru}
H_{\rm ext}(t)=-\left(d_0 q+\sum_{S=L,R}\,\sum_{k=1}^\infty d_k \,q_{S, k}\right) F(t) \, ,
\end{equation}
with a coupling constant $d_0$ to the central particle and a coupling constant $d_k$ to the $k^{\rm th}$ particle of each chain. After the transformation given in Eq.\ (\ref{trans}), this becomes
\begin{eqnarray}\nonumber
H_{\rm ext}(t)&=&-\left(d_0 q+\sum_{S=L,R}\int_0^\pi d\psi \,d(\psi) x_S (\psi)\right) F(t) \\ 
&=&-\left(d_0 q+\int_{-\pi}^\pi d\psi \,d(\psi) x(\psi)\right) F(t) \, ,
\end{eqnarray}
where
\begin{equation}\label{dpsi}
d(\psi)= \sqrt{\frac{2}{\pi}}\,\sum_{k=1}^\infty d_k \sin(k\vert\psi\vert) \, .
\end{equation}
This form corresponds to the coupling term (\ref{Hext}) of the extended Caldeira-Leggett model.
Hence, the driven Rubin model can serve as a concrete example for the general theory outlined in Sec.~\ref{sec:two}.

\subsection{Harmonic potential and finite range force coupling}
To illustrate the theory, we consider specifically a Rubin model where the central particle moves in a harmonic potential
\begin{equation}\label{ham}
V(q)=\frac{1}{2}M\omega_0^2 q^2 \, .
\end{equation}
Furthermore, we assume that the coupling coefficients $d_k$ decrease exponentially with the distance form the center. Specifically,
\begin{equation}\label{dkrho}
d_k=d_c\,e^{-\rho k} \, ,
\end{equation}
where $1/\rho$ is a measure for the range of the force field. We then obtain from Eq.~(\ref{dpsi})
 \begin{equation}\label{dpsi2}
d(\psi)
=\sqrt{\frac{1}{2\pi}}\,\frac{d_c \vert \sin(\psi)\vert}{\cosh(\rho)-\cos(\psi)} \, .
\end{equation}
The force delay kernel (\ref{lambda}), which characterizes the influence of the driving force mediated by the bath, takes the form
\begin{eqnarray}\label{lambdaR}
\lambda(t)&=&\int_{-\pi}^\pi d\psi\, \frac{c(\psi) d(\psi)}{m\omega(\psi)} \sin\left( \omega(\psi) t\right)  \\ \nonumber
&=&\frac{2Kd_c}{\pi}\int_{0}^\pi d\psi\, \frac{\sin^2(\psi) \sin\left( \sin(\psi/2) \omega_D t\right)}{m\omega_D\sin(\psi/2)[\cosh(\rho)-\cos(\psi)]}   \\ \nonumber
&=&\frac{4\omega_D d_c}{\pi}\int_{0}^1 dx\, \frac{x\sqrt{1-x^2} \sin\left( x \omega_D t\right)}{\cosh(\rho)-1+2x^2} \, . 
\end{eqnarray}
For later purposes we consider the Laplace transform
\begin{equation}
\hat\lambda(z)=\frac{4d_c}{\pi}\int_{0}^1 dx\, \frac{x^2\sqrt{1-x^2} }{[\cosh(\rho)-1+2x^2][(z/\omega_D)^2+x^2]} \, ,
\end{equation}
which is a function of the dimensionless variable $\nu=z/\omega_D$. The integral can be evaluated with the result
\begin{equation}
\hat\lambda(z)=d_c\,\frac{\nu\sqrt{1+\nu^2}-\nu^2-a\sqrt{1+a^2}+a^2}{\nu^2-a^2} \, ,
\end{equation}
where
\begin{equation}
a^2=\frac{\cosh(\rho)-1}{2} \, .
\end{equation}

\subsection{Response function}
\begin{figure}
\includegraphics[width=0.36\textwidth]{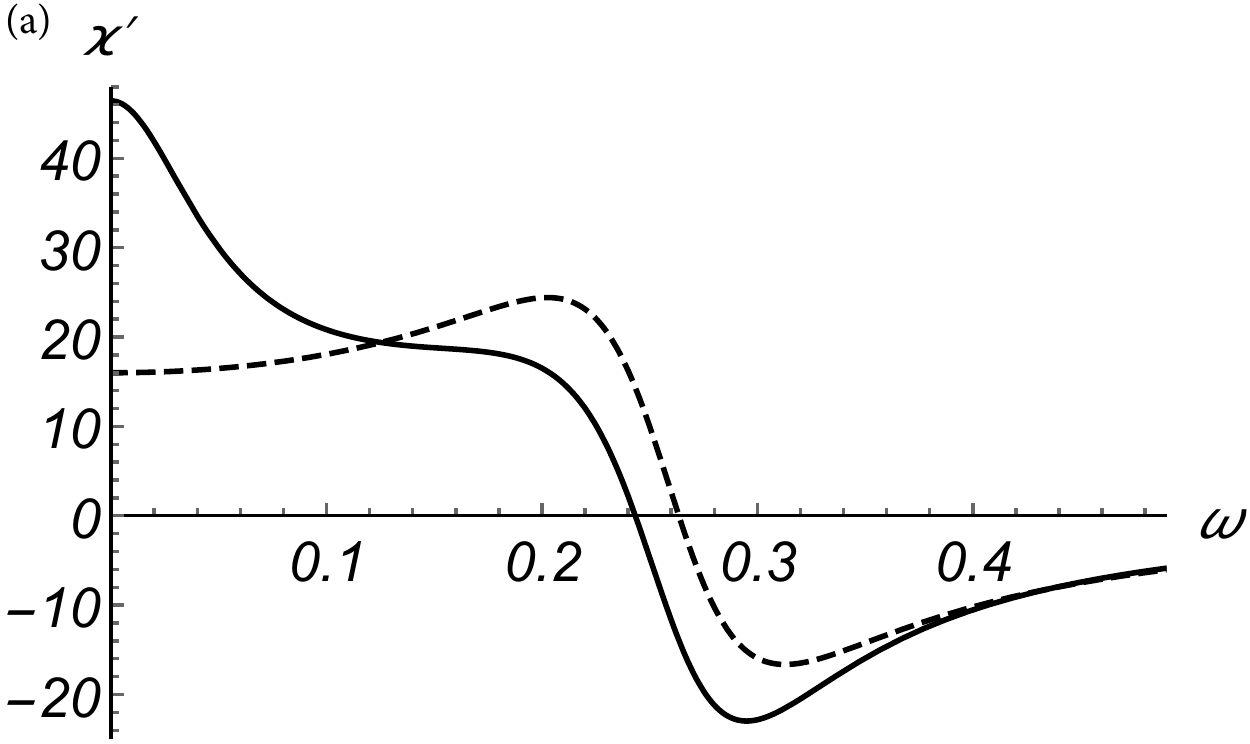} \\
\vspace{0.3cm}
\includegraphics[width=0.35\textwidth]{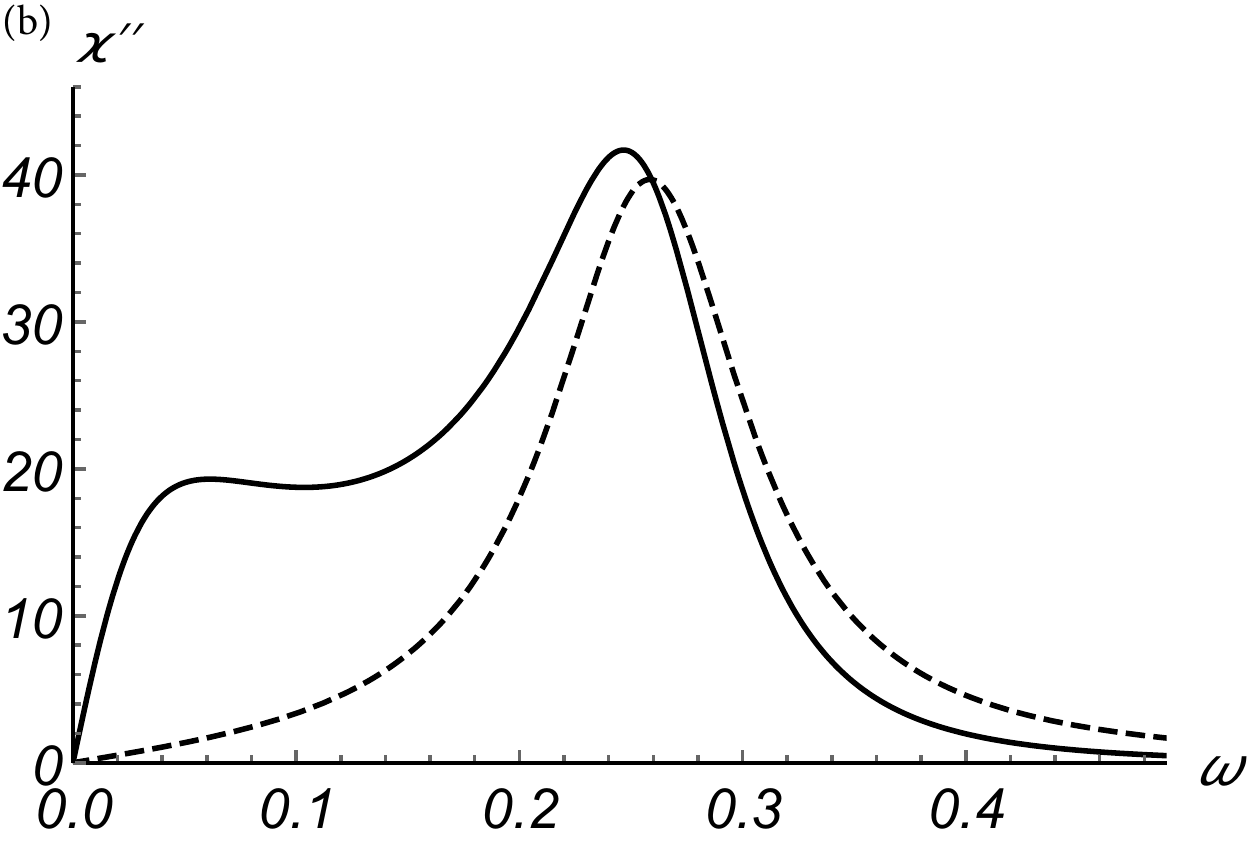}
\caption{\label{fig2} The dynamic susceptibility $\chi=\chi'+i\chi''$ in units of $d_0/M\omega_D^2$ is shown as a function of the driving frequency $\omega$ in units of $\omega_D$ for a Rubin model with mass ratio $M=10\,m$ and an oscillation frequency $\omega_0=0.25\,\omega_D$. The upper [lower] panel shows the real [imaginary] part (or, the dispersive [absorptive] part) of $\chi(\omega)$. The dashed lines give results for $d_c=0$ when only the central particle is driven by the applied force. The full lines include the driving of the bath chains for coupling constant $d_c=0.1\,d_0$ and range parameter $\rho=0.1$.}
\end{figure}

Next, we study the response of the average position $\langle q(t)\rangle$ of the central particle to a harmonic applied force 
\begin{equation}\label{FOm}
F(t)=F\cos(\omega t) \, .
\end{equation}
From the equation of motion (\ref{le}), we obtain for the harmonic potential (\ref{ham})
\begin{eqnarray}\label{le2}\nonumber 
&&M\langle \ddot q(t)\rangle + M\int_0^t ds \, \gamma(t-s)\langle \dot q(s)\rangle 
+M\omega_0^2 \langle q(t)\rangle\\
&&\quad = d_0 F(t) +\int_0^t ds\, \lambda(t-s)F(s) \, .
\end{eqnarray}
In the long time limit, this yields
\begin{equation}
\langle q(t)\rangle = \hbox{Re}\left\{\frac{1}{M}\frac{d_0 +\hat \lambda(-i\omega)  }{\omega_0^2-i\omega\hat\gamma(-i\omega)-\omega^2}\,F\,e^{-i\omega t}\right\} \, .
\end{equation}
From this result, we extract the dynamic susceptibility 
\begin{equation}
\chi(\omega)=\frac{1}{M}\frac{d_0 +\hat \lambda(-i\omega)  }{\omega_0^2-i\omega\hat\gamma(-i\omega)-\omega^2} \, .
\end{equation}
The influence of the driven bath modes is described by the function $\hat\lambda(-i\omega)$ in the numerator. 
In Fig.~\ref{fig2}, we show the real (dispersive) and imaginary (absorptive) parts of the dynamic susceptibility for $M=10\,m, \omega_0=0.25\,\omega_D, d_c=0.1\,d_0$ and $\rho=0.1$. We compare the results for conventional driving ($d_c=0$), when the driving field couples to the central mass only, with the case when the driven bath modes are taken into account.

We see that the response to a driven bath acquires additional nontrivial features. In the low-frequency sector the response of the dispersive part is strongly enhanced by the bath driving. A maximum develops at zero frequency. Likewise, the response of the absorptive part is also enhanced in the low-frequency sector. An additional shoulder emerges due to the bath driving. In general, both response curves experience a slight shift in frequency towards smaller frequencies, as compared to the case without bath driving. This opens a way to manipulate and control the response characteristics of the central system by driving the bath modes. 


\section{Driven two-state system}\label{sec:four} 
As another example, we consider a damped two-state system coupled to a bath of harmonic oscillators and subject to external driving. This spin-boson model \cite{Leggett1987,Weiss2008} is a hallmark model to study the influence of quantum dissipation and dephasing on coherent quantum transitions and has found extensive applications in physics, chemistry and biology. In contrast to earlier studies \cite{GrifoniDrivenTunneling}, we consider the case when the external driving force excites not only the two-state system directly, but also the bath oscillators.  

\subsection{Model Hamiltonian}

The microscopic Hamiltonian of the model may be written as
\begin{equation}\label{Htot}
H = H_{SB} + H_{\rm ext}(t)\, ,
\end{equation}
where 
\begin{equation}\label{HSB}
H_{SB} = H_S + H_R + H_{\rm int} 
\end{equation}
is the familiar spin-boson Hamiltonian in which
\begin{equation}\label{Hs}
H_S = \tfrac{1}{2}\epsilon \sigma_z - \tfrac{1}{2}\Delta\sigma_x
\end{equation}
is the Hamiltonian of a two-state system described in terms of the Pauli matrices $\sigma_i \, (i=x,y,z)$. Here $\epsilon$ is the energy difference between the two eigenstates of $\sigma_z$ with eigenvalues $+1$ and $-1$, and the second term of Eq.~(\ref{Hs}) describes tunneling between these two states. $H_R$ is again the Hamiltonian (\ref{HR}) of a bath of harmonic oscillators, and
\begin{equation}\label{HintSB}
H_{\rm int} = -\tfrac{1}{2}\sigma_z \sum_n c_n x_n
\end{equation}
is the coupling between system and bath. The standard spin-boson model (\ref{HSB}) describes a two-state system coupled to an environment of harmonic oscillators with the effective spectral density (\ref{Jeff}).
We extend the conventional spin-boson Hamiltonian by the additional term
\begin{equation}\label{HextSB}
H_{\rm ext}(t) = -\left( \tfrac{1}{2} d_0 \sigma_z + \sum_n d_n x_n\right) F(t)
\end{equation}
in Eq.~(\ref{Htot}) which adds the coupling to an external driving force $F(t)$. We have allowed for a direct coupling of the external field to the two-state system with coupling constant $d_0$ as well as a coupling to the bath oscillators with coupling constants $d_n$. We remark that the coefficients $c_n$ and $d_0$ introduced here have physical dimensions different from the dimensions of the corresponding coefficients of the potential model discussed in Section \ref{sec:two}. This implies also different physical dimensions of other quantities such as the spectral density $J(\omega)$ or the force delay kernel $\lambda(t)$.

\subsection{Evolution equations}

The Heisenberg equations of motion of the bath oscillators
\begin{eqnarray}\label{Heomx2}
\dot x_n &=& \frac{p_n}{m_n}
\\\label{Heomp2}
\dot p_n &=& -m_n\omega_n^2 x_n + \tfrac{1}{2} c_n \sigma_z + d_n F
\end{eqnarray}
are solved by
\begin{eqnarray}\label{xnoft}
x_n(t) &=& x_{n0}\cos\left(\omega_n t\right) + \frac{p_{n0}}{m_n\omega_n}\sin\left(\omega_n t\right)
\\
& \nonumber +&\frac{1}{m_n\omega_n} \int_0^t ds \sin\left[ \omega_n(t-s)\right] \left[ d_n F(s) + \tfrac{1}{2} c_n \sigma_z(s) \right] \, ,
\end{eqnarray}
where $x_{n0}$ and $p_{n0}$ are the Heisenberg operators at time $t=0$.

On the other hand, the Heisenberg equations of motion of the two-state system read
\begin{eqnarray}\label{eomsigmax}
\dot \sigma_x &=& -\frac{1}{\hbar} \left(\epsilon - \sum_n c_n x_n - d_0 F \right) \sigma_y
\\\label{eomsigmay}
\dot \sigma_y &=& \frac{1}{\hbar} \left(\epsilon - \sum_n c_n x_n - d_0 F \right) \sigma_x +\frac{\Delta}{\hbar}\, \sigma_z 
\\\label{eomsigmaz}
\dot \sigma_z &=& -\frac{\Delta}{\hbar}\,  \sigma_y \, .
\end{eqnarray}
Now, from Eq.~(\ref{xnoft}) we find 
\begin{eqnarray}\label{cnxn}
\sum_n c_n x_n(t) &=& \int_0^t ds\, \zeta(t-s) \sigma_z(s) \nonumber \\
&&+ \xi(t) +\int_0^t ds \,\lambda(t-s) F(s) \, .
\end{eqnarray}
Here, we have introduced the kernel 
\begin{equation}\label{chiSB}
\zeta(t) = \sum_n \frac{c_n^2}{2 m_n\omega_n}  \sin\left( \omega_n t\right)
\end{equation}
describing the temporal response of the bath. It can be written in terms of the effective spectral density of bath modes (\ref{Jeff}) as
\begin{equation}\label{chiJ}
\zeta(t)=\int_0^{\infty}\frac{d\omega}{\pi}J(\omega)\,\sin(\omega t) \, .
\end{equation}
The random force operator
\begin{equation}\label{xi}
\xi(t) =\sum_n c_n \left[x_{n0}\cos\left(\omega_n t\right) + \frac{p_{n0}}{m_n\omega_n}\sin\left(\omega_n t\right)\right]
\end{equation}
depends on the initial values of the bath operators $x_n, p_n$. When the bath is initially equilibrated at inverse temperature $\beta=1/(k_B T)$, the noise force is Gaussian with vanishing mean value 
\begin{equation}
\langle \xi(t)\rangle =0
\end{equation}
and the noise correlation function 
\begin{eqnarray}\nonumber\label{xixi}
\langle \xi(t+s)\xi(s)\rangle &=&C(t) =	 \sum_n \frac{\hbar c_n^2}{2m_n\omega_n} \bigg[\coth\left(\frac{\beta\hbar\omega_n}{2}\right)\\ 
&&\times\cos(\omega_n t)  -i\sin(\omega_nt)\bigg] \, .
\end{eqnarray}
In terms of the spectral density, this correlator is given by
\begin{equation}\label{Coft}
C(t)=\hbar\int_0^{\infty}\!\frac{d\omega}{\pi}J(\omega)\bigg[\coth\left(\frac{\beta\hbar\omega}{2}\right)\cos(\omega t)  -i\sin(\omega t)\bigg] \, .
\end{equation}

The last term in Eq.~(\ref{cnxn}) describes the time-retarded effect of the applied force mediated by the driven bath modes. This effect is again characterized by the force delay kernel $\lambda(t)$ introduced in Eq.~(\ref{lambda}).
The evolution equations (\ref{eomsigmax}) - (\ref{eomsigmaz}) depend on the applied force only via the term 
\begin{eqnarray}\nonumber
&&\sum_n c_nx_n(t)+d_0F(t)=\int_0^t ds\, \zeta(t-s) \sigma_z(s)+ \xi(t) \\ 
&&\qquad+d_0F(t) +\int_0^t ds \,\lambda(t-s) F(s) \, ,
\end{eqnarray}
where the last two terms arise from the applied force.  Hence, the driving of the bath modes effectively adds to the direct force $d_0 F(t)$ a retarded force $\int_0^t ds\, \lambda(t-s)F(s)$. Accordingly, the dynamics of a two-state system in a driven bath can be described by the standard spin-boson model driven by an effective force of the from (\ref{Feff}).
With this observation, the entire formalism for the dissipative quantum-mechanical two-state system can be extended to externally driven baths. 

\subsection{Weak damping and weak driving}
Rather than employing powerful path integral techniques to the dissipative two-state system \cite{Leggett1987,Weiss2008}, we shall present here an approach based on the Bloch-Langevin type equations of motion derived above.  Since our aim is to illustrate the effect of a driven bath, we shall restrict ourselves to a symmetric two-state system. The evolution equations  (\ref{eomsigmax}) -- (\ref{eomsigmaz}) and the relation (\ref{cnxn}) give for the time rates of change of the Pauli matrices (for $\epsilon=0$) 
\begin{eqnarray}\nonumber\label{eomx}
\hbar   \dot \sigma_x(t)   &=&  \int_0^t ds\, \zeta(t-s)   \sigma_z(s)\,\sigma_y(t)  \\ 
&&\qquad +[\xi(t)+ F_{\rm eff}(t)]    \sigma_y(t)  \, ,
\\\label{eomy}\nonumber
\hbar \dot \sigma_y(t)   &=&\Delta\,  \sigma_z (t)   - \int_0^t ds\, \zeta(t-s)   \sigma_z(s)\,\sigma_x(t) \\ 
&&\qquad - [\xi(t)+ F_{\rm eff}(t)]   \sigma_x(t)  \, ,
\\ \label{eomz}
\hbar \dot \sigma_z(t)   &=& -\Delta\,   \sigma_y(t)  \, ,
\end{eqnarray}
where we have used the effective force (\ref{Feff}). Clearly, the spin values couple to two-point products. This leads ultimately to a hierarchy of equations of motion which can be truncated for weak damping. Furthermore, we shall restrict ourselves to the linear response of the spin.

The derivation of the evolution equations in the limit of weak damping and driving is presented in Appendix~\ref{app:A}. In this limit the dynamics is characterized by the Larmor frequency
\begin{equation}\label{Lf}
\omega_0=\frac{\Delta}{\hbar}\, .
\end{equation}
and damping coefficients 
\begin{equation}\label{gacr}
\gamma_c^{\prime}=  \frac{J(\omega_0)}{2\hbar}\coth\left(\frac{\beta\hbar\omega_0}{2}\right)\, ,
\end{equation}
\begin{equation}\label{gaci}
\gamma_c^{\prime\prime}= -\frac{1}{\hbar} P\int_0^{\infty}\frac{d\omega}{\pi}J(\omega)\, \frac{\omega}{\omega^2-\omega_0^2}\, ,
\end{equation}
\begin{equation}\label{gasr}
\gamma_s^{\prime} =- \frac{1}{\hbar}P\int_0^{\infty}\frac{d\omega}{\pi}J(\omega)\coth\left(\frac{\beta\hbar\omega}{2}\right)\, \frac{\omega_0}{\omega^2-\omega_0^2}\, ,
\end{equation}
and
\begin{equation}\label{gasi}
\gamma_s^{\prime\prime}= -\frac{J(\omega_0)}{2\hbar}\, ,
\end{equation}
where $P$ denotes the Cauchy principal value.

As shown in Appendix~\ref{app:A}, in terms of these parameters the evolution equations for the average spin components take the form
\begin{eqnarray}\nonumber\label{eomav}
 \langle \dot \sigma_x(t) \rangle   &=&   -\gamma_s''   -\gamma_c'\langle\sigma_x(t)\rangle  +\frac{1}{\hbar} F_{\rm eff}(t)\langle \sigma_y(t) \rangle  \, ,
\\\nonumber
\langle \dot \sigma_y(t) \rangle  &=&(\omega_0+\gamma_s')\, \langle  \sigma_z (t) \rangle   - \gamma_c' \langle  \sigma_y(t)\rangle  -\frac{1}{\hbar} F_{\rm eff}(t)  \langle \sigma_x(t) \rangle\, ,
\\
\langle \dot \sigma_z(t)\rangle   &=& -\omega_0\,  \langle \sigma_y(t) \rangle  \, .
\end{eqnarray}

Let us first consider the steady state in the absence of driving, i.e., for $F_{\rm eff}(t)=0$.
The conditions $\langle\dot\sigma_i\rangle_{\rm eq}=0$ for $i=x,y,z$ yield 
\begin{equation}
\langle \sigma_y\rangle_{\rm eq}= \langle  \sigma_z  \rangle_{\rm eq} =0
\end{equation}
and
\begin{equation}\label{sigxeq}
\langle \sigma_x\rangle_{\rm eq}=-\frac{\gamma_s''}{\gamma_c'}=\tanh\left(\frac{\beta\hbar\omega_0}{2}\right)\, ,
\end{equation}
where the second relation follows from Eqs.~(\ref{gacr}) and (\ref{gasi}). This is the canonical expectation value of a spin-1/2 with Hamiltonian $H_S=-\tfrac{1}{2}\hbar\omega_0\sigma_x$ in thermal equilibrium with a bath at inverse temperature $\beta$.

\subsection{\boldmath  Linear response of $\langle \sigma_z(t)\rangle$}
With the equations of motion for the expectation values of the spin components at hand, we may study next the linear response of $\langle\sigma_z(t)\rangle$ to an applied force $F(t)$. From the Eqs.~(\ref{eomav}), we get
\begin{eqnarray}\label{eomy2}
&&\langle \ddot \sigma_y(t) \rangle +\gamma_c' \langle \dot \sigma_y(t)\rangle +(\omega_0+\gamma_s')\,\omega_0\, \langle  \sigma_y (t) \rangle \, , \\ \nonumber
&&\qquad\quad =-\frac{1}{\hbar} \left[ \dot F_{\rm eff}(t)\langle\sigma_x(t)\rangle+F_{\rm eff}(t)\langle\dot\sigma_x(t)\rangle\right]\, , 
\end{eqnarray}
which in the long time limit has the formal solution 
\begin{eqnarray}\nonumber\label{rey}
\langle \sigma_y(t) \rangle&=& 
-\frac{1}{\hbar}\int_0^\infty ds\, e^{-\tfrac{1}{2}\gamma_c's}\left[\cos(\omega_rs)-\frac{\gamma_c'}{2\omega_r}\sin(\omega_rs)\right] \\
&&\qquad\quad\times
 F_{\rm eff}(t-s)\langle\sigma_x(t-s)\rangle \, .
\end{eqnarray}
Here, we have introduced
\begin{equation}\label{omr}
\omega_{r}=\sqrt{\omega_0^2+\gamma_s'\omega_0-\gamma_c^{\prime\,2}/4} \, .
\end{equation}
To linear order in the applied force, we can replace $\langle\sigma_x(t-s)\rangle$ on the left hand side of Eq.~(\ref{rey})	by its equilibrium value $\langle\sigma_x\rangle_{\rm eq}$ given in Eq.\ (\ref{sigxeq}). Furthermore, we shall assume driving by the sinusoidal force  (\ref{FOm}) which in the long time limit implies an effective force of the form
\begin{equation}
F_{\rm eff}(t)=F\,\hbox{Re}\left\{\big[d_0+\hat\lambda(-i\omega)\big]e^{-i\omega t}\right\} \, .
\end{equation}
Inserting these relations into Eq.~(\ref{rey}), the linear response of $\langle\sigma_y(t)\rangle$ is obtained as
\begin{equation}
\langle \sigma_y(t) \rangle= 
\frac{\langle\sigma_x\rangle_{\rm eq}}{\hbar}\,\hbox{Re}\left\{\frac{i\omega\big[d_0+\hat\lambda(-i\omega)\big]}{\omega_r^2+\gamma_c^{\prime\,2}/4-i\gamma_c'\omega-\omega^2} F\,e^{-i\omega t} \right\} \, .
\end{equation}
Since $\langle\dot\sigma_z(t)\rangle=-\omega_0\langle\sigma_y(t)\rangle$,  we also get straightforwardly the linear response of $\langle\sigma_z(t)\rangle$ as
\begin{equation}
\langle \sigma_z(t) \rangle= 
\frac{\langle\sigma_x\rangle_{\rm eq}}{\hbar}\,\hbox{Re}\left\{\frac{\omega_0\big[d_0+\hat\lambda(-i\omega)\big]}{\omega_r^2+\gamma_c^{\prime\,2}/4-i\gamma_c'\omega-\omega^2} F\,e^{-i\omega t}\right\} \, .
\end{equation}
From this result we extract the dynamic susceptibility
\begin{equation}\label{chiz}
\chi_z(\omega)=\frac{\langle\sigma_x\rangle_{\rm eq}}{\hbar}\,\frac{\omega_0\big[d_0+\hat\lambda(-i\omega)\big]}{\omega_0(\omega_0+\gamma_s') -\omega(\omega+i\gamma_c')} \, ,
\end{equation}
where we have employed Eq.~(\ref{omr}).

\section{Two-state system coupled to harmonic chains}\label{sec:five}
As a concrete example for a dissipative two-state system we consider a modified Rubin model with a central symmetric two-state system. 
The Hamiltonian is taken as 
\begin{eqnarray}\label{Hrubi}
H_{\rm TS }&=&-\frac{1}{2}\Delta\sigma_x +\sum_{S=L,R}\Bigg\{\frac{K}{2}[q_{S, 1}^2-\breve{q}q_{S,1}\sigma_z] \qquad \\ \nonumber
&&+\sum_{k=1}^\infty\left[ \frac{p_{S, k}^2}{2m}+\frac{K}{2}\left(q_{S, k+1}-q_{S, k}\right)^2\right]\Bigg\} \, ,
\end{eqnarray}
As in the Rubin model (\ref{Hru}), the Hamiltonian (\ref{Hrubi}) describes two semi-infinite harmonic chains coupled to a central object which is now a two-state system. Here, the particles with coordinates $q_{L,1}$ and $q_{R,1}$ at the ends of the chains  couple to $\breve{q}\sigma_z$. When the two-state system is a tunneling system with tunnel splitting $\Delta$, the parameter $\breve{q}$ describes the tunneling length.

One can again employ the transformation (\ref{trans}) which gives
\begin{eqnarray}\label{Hrubi2}
\nonumber
H_{\rm TS}&=&-\frac{1}{2}\Delta\sigma_x +\int_{-\pi}^\pi d\psi\, \Bigg[ \frac{p(\psi)^2}{2m}  
+\frac{1}{2}m\omega(\psi)^2 x(\psi)^2  \\  
&&\qquad -\sqrt{\frac{1}{2\pi}}K \breve{q}\vert \sin(\psi)\vert  x(\psi)\sigma_z\Bigg] \, ,
\end{eqnarray}
with the bath mode frequencies (\ref{om}).
This is a symmetric spin-boson model of the form of Eq.\ (\ref{HSB}). The last term is the interaction term of Eq.\ (\ref{HintSB}) with coupling constants
\begin{equation}\label{cTR}
c_{\rm TS}(\psi)=\breve{q}\, c(\psi) \, ,
\end{equation}
where $c(\psi)$ is given in Eq.~(\ref{cpsi}).

We now add an external force $F(t)$ coupling to the two-level system and to the bath oscillators. The coupling Hamiltonian is assumed to be again of the form (\ref{Hextru}), with $q$ replaced by $\frac{1}{2}\breve{q}\sigma_z$. Accordingly, we obtain in the transformed variables
\begin{equation}\label{HextSB2}
H_{\rm ext}(t)=-\left(\tfrac{1}{2} d_0 \breve{q}\sigma_z+\int_{-\pi}^\pi d\psi \,d(\psi) x(\psi)\right) F(t) \, ,
\end{equation}
where $d(\psi)$ is again given by Eq.~(\ref{dpsi}). The driving Hamiltonian (\ref{HextSB2}) is indeed of the form (\ref{HextSB}) with the coupling constant
\begin{equation}\label{dTR}
d_{0,{\rm TS}}=\breve{q}\, d_0 \, .
\end{equation}

In view of relation (\ref{cTR}), the effective spectral density of the modified two-state Rubin model is
\begin{equation}\label{JTR}
J_{\rm TS}(\omega)= \breve{q}^2 J(\omega) \, ,
\end{equation}
where $J(\omega)$ is the spectral density (\ref{JR}) of the standard Rubin model. Furthermore, assuming again coupling coefficients of the applied force to the reservoir particles of the form (\ref{dkrho}), the force delay kernel is
\begin{equation}
\lambda_{\rm TS}(t)=\breve q \,\lambda(t) \, ,
\end{equation}
where $\lambda(t)$ is the kernel (\ref{lambdaR}). Using Eqs.~(\ref{JR}) and (\ref{JTR}), we obtain from Eqs.~(\ref{gacr}) and (\ref{gasr}) the damping coefficients
\begin{equation}
\gamma_c'=\frac{m\omega_0\breve{q}^2\sqrt{\omega_D^2-\omega_0^2}}{2\hbar} \coth\left(\frac{\beta\hbar\omega_0}{2}\right)
\end{equation}
and
\begin{equation}
\gamma_s^{\prime} = \frac{m\omega_0\breve{q}^2}{\hbar}P\int_0^{\omega_D}\frac{d\omega}{\pi}\frac{\omega\sqrt{\omega_D^2-\omega^2}}{\omega_0^2-\omega^2}\coth\left(\frac{\beta\hbar\omega}{2}\right) \, .
\end{equation}
\begin{figure}
\includegraphics[width=0.36\textwidth]{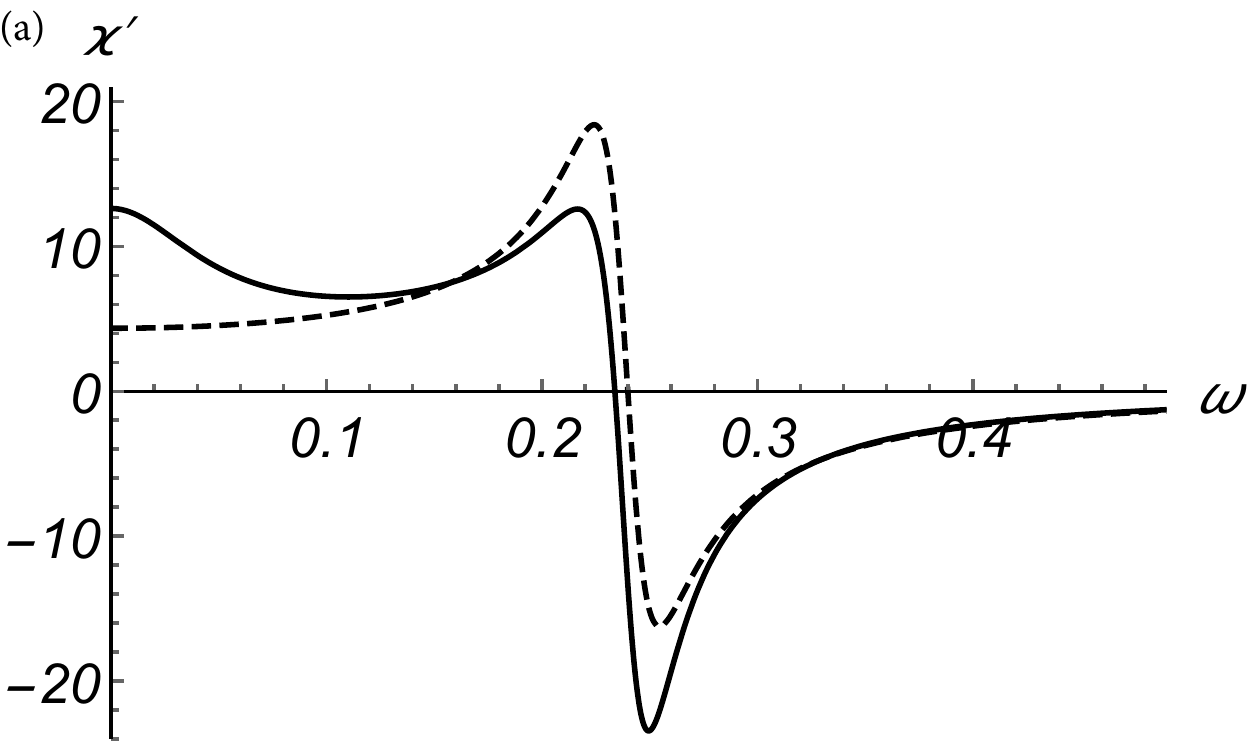}\\
\vspace{0.3cm}
\ \includegraphics[width=0.35\textwidth]{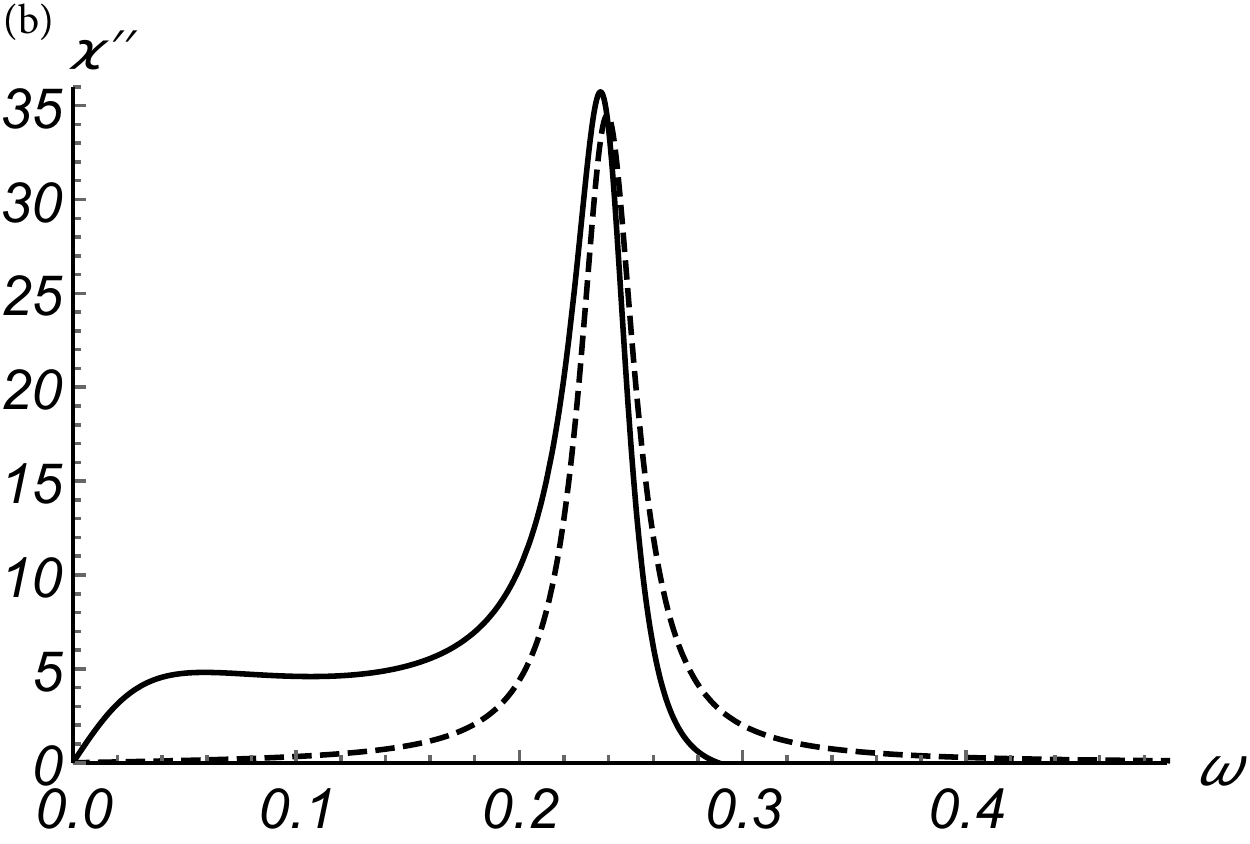}
\caption{\label{fig3} The dynamic susceptibility $\chi_z=\chi'_z+i\chi''_z$ in units of $d_{0,{\rm TS}}/\hbar\omega_D$ is shown as a function of the driving frequency $\omega$ in units of $\omega_D$ for a modified Rubin model with central two-state system. The Larmor frequency $\omega_0=0.25\,\omega_D$,  the tunneling length  $\breve{q}=0.5\, \sqrt{\hbar/m\omega_D}$ and the temperature $k_BT=0$.  The upper [lower] panels show the real [imaginary] part of $\chi(\omega)$. The dashed lines give results for $d_c=0$ when only the two-state system is driven by the applied force. The full lines include the driving of the reservoir particles for coupling constant $d_c=0.1\,d_{0,{\rm TS}}/\breve q$ and range parameter $\rho=0.1$.}
\end{figure}
\begin{figure}
\includegraphics[width=0.36\textwidth]{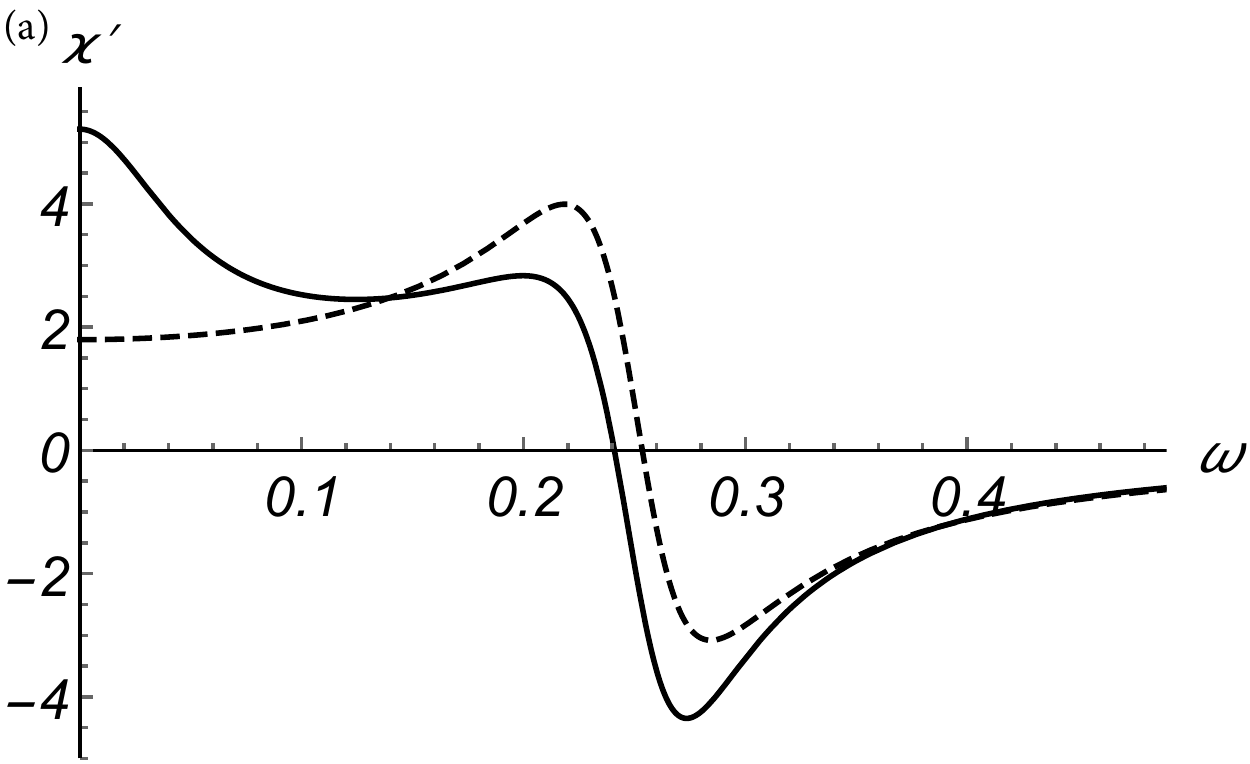}\\
\vspace{0.3cm}
\includegraphics[width=0.35\textwidth]{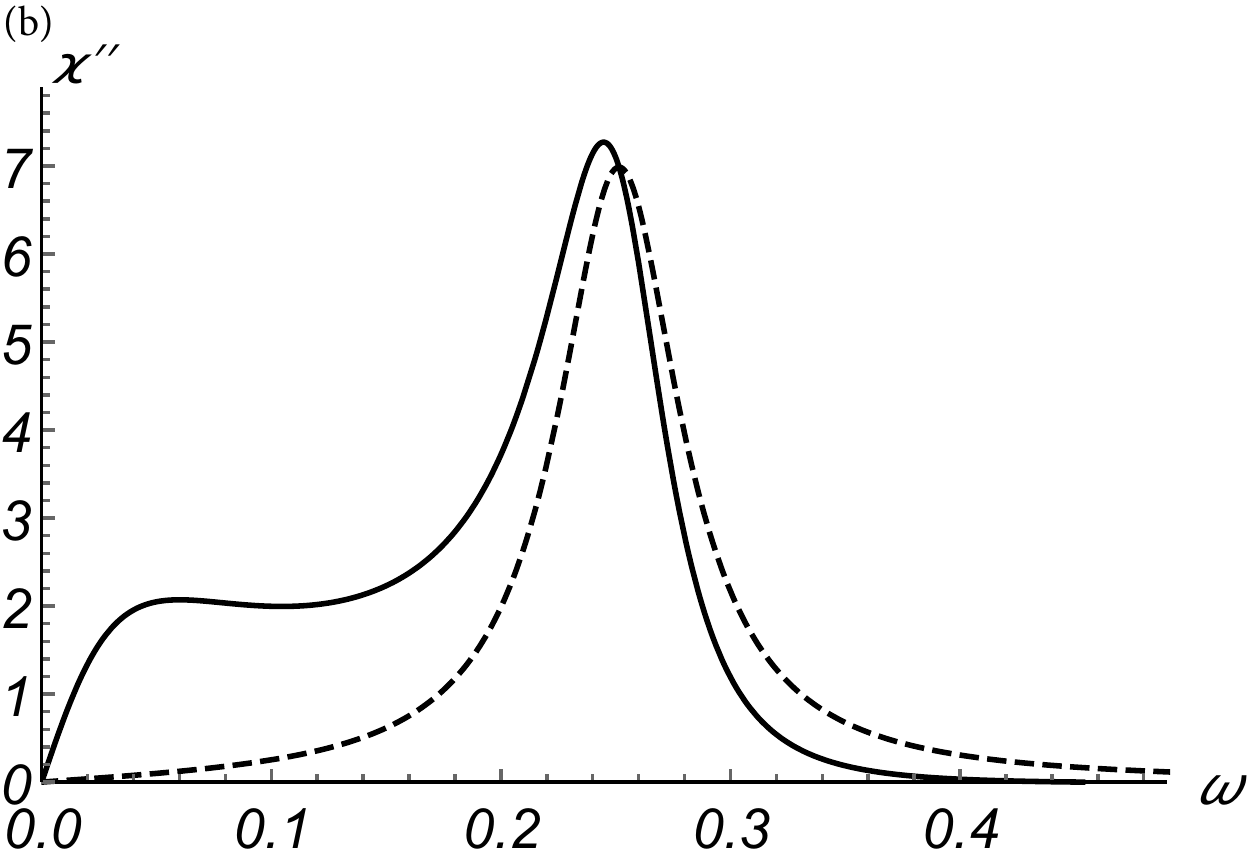}
\caption{\label{fig4} Same as Fig.~\ref{fig3} for finite temperature $k_BT=0.25\,\hbar\omega_D$.}
\end{figure}
With these results at hand, we can now determine the response function (\ref{chiz}).
Figs.~\ref{fig3} and \ref{fig4} show the dynamic susceptibility $\chi_z(\omega)$ for a particular set of parameters at zero and finite temperatures. In contrast to the case of Brownian motion in a harmonic potential, the dynamic susceptibility now depends on the temperature. The effect of the driven bath modes, however, is qualitatively similar leading to an enhanced response for frequencies below the resonance frequency $\omega_0$. The dispersive parts develop a maximum at zero frequency, while the absorptive parts acquire a shoulder-like feature at low frequencies.

\section{Conclusions}\label{sec:six}
We have discussed the response characteristics of a quantum system immersed in a Caldeira-Leggett bath whose degrees of freedom are explicitly driven by an external time-dependent force, in addition to the direct time-dependent forcing of the system itself. As a concrete example, we have considered the Rubin model of a chain of quantum particles coupled by linear springs. The particles have equal mass apart from a ''central'' particle which has a larger mass.  We have shown that in the presence of time-dependent driving this model can be mapped to a quantum system which couples to a time-dependent Caldeira-Leggett bath. The effect of the bath driving was shown to be fully captured by a time-dependent force whose momentary value depends on the prehistory of the system-plus-bath giving rise to non-Markovian retardation effects. 

We then have chosen two specific examples, a central particle moving in a harmonic potential and a central two-state system. In the case of the harmonic potential, we have solved the quantum dynamics of the central system in the driven Caldeira-Leggett bath and have determined its full response. In the case of a quantum two-state system, we have assumed a weak system-bath coupling for all concrete results and have iterated the time evolution equations up to the lowest non-trivial order in the system-bath coupling. This has allowed us to evaluate the dynamical response of the quantum two-state system to a periodic driving of the system and the bath. 

The dynamic susceptibility was shown to be altered qualitatively by the bath drive: The dispersive (or, real) part of the susceptibility is enhanced at low frequencies and acquires a maximum at zero frequency. The absorptive (or, imaginary) part is also enhanced in the low-frequency regime and develops a shoulder-like behavior there. These features seem to be generic and were found for the harmonic potential and for the quantum two-state system. 

Our results illustrates that the bath can be used to modify the characteristics of the frequency dependent  response of the central system. Although a single-frequency driving mode is applied, the response of the system is finite in a broad frequency band. In that sense, the driven bath acts as a dispersive frequency modulator.  

The methods presented here can straightforwardly be extended to other, potentially more complex dissipative quantum systems in a driven bath. The basic assumptions underlying the theory are, firstly, that the response of the bath to the system dynamics can be described by means of linear response theory. This characterizes linear dissipation and allows for a description within the Caldeira-Leggett approach. Secondly, the applied force exciting the bath modes directly should not drive them beyond the range of validity of linear response theory. Whenever these conditions are satisfied an effective time-retarded driving force can be defined and the basic features of the theory will persist. 

\begin{acknowledgments}
This work was supported by the Excellence Cluster "The Hamburg Center for Ultrafast Imaging - Structure, Dynamics and Control of Matter at the Atomic Scale" of the Deutsche Forschungsgemeinschaft.
\end{acknowledgments}

\appendix
\section{Weak damping limit for weakly driven two-state system}\label{app:A}

In the absence of damping and driving the evolution equations (\ref{eomx}) -- (\ref{eomz})  give
\begin{eqnarray}\label{evo0}
\sigma_x(s+u) &=&\sigma_x(s)\, ,\\ \nonumber
\sigma_y(s+u) &=&\sigma_y(s)\cos\left(\omega_0 u \right)+ \sigma_z(s)\sin\left(\omega_0u \right)\, ,\\ \nonumber
\sigma_z(s+u) &=& \sigma_z(s)\cos\left(\omega_0u \right)- \sigma_y(s)\sin\left(\omega_0u \right)\, ,
\end{eqnarray}
where $\omega_0$ is the Larmor frequency (\ref{Lf}). The Eqs.~(\ref{evo0}) imply
\begin{equation}
\sigma_z(s) =\sigma_z(t) \cos\boldsymbol{\left(}\omega_0(t-s) \boldsymbol{\right)} + \sigma_y(t)\sin\boldsymbol{\left(}\omega_0(t-s) \boldsymbol{\right)} \, .
\end{equation}
For weak damping this result can be inserted into the retardation terms of the evolution equations (\ref{eomx}) and (\ref{eomy}). In the long time limit, we then obtain
\begin{eqnarray}\nonumber\label{eom2}
\hbar   \dot \sigma_x(t)   &=&  \hbar\kappa_s -i\hbar\kappa_c \sigma_x(t)  + [\xi(t)+ F_{\rm eff}(t)]   \sigma_y(t)  \, ,
\\\nonumber
\hbar \dot \sigma_y(t)   &=&\hbar\omega_0\,  \sigma_z (t)  - i\hbar\kappa_c\sigma_y(t)+i\hbar\kappa_s \sigma_z(t)  \\ \nonumber
&&\qquad - [\xi(t)+ F_{\rm eff}(t)]  \sigma_x(t)  \, ,
\\
\hbar \dot \sigma_z(t)   &=& -\hbar\omega_0\,   \sigma_y(t)  \, ,
\end{eqnarray}
where
\begin{equation}
\kappa_c=\frac{1}{\hbar}\int_0^\infty dt\, \zeta(t)\cos\left(\omega_0t \right)
\end{equation}
and
\begin{equation}
\kappa_s=\frac{1}{\hbar}\int_0^\infty dt\, \zeta(t)\sin\left(\omega_0t \right) \, .
\end{equation}
Furthermore, we have made use of the multiplication rules for Pauli matrices.
Using the form (\ref{chiJ}) of the kernel $\zeta(t)$, we find
\begin{equation}\label{kac}
\kappa_c=\frac{1}{\hbar}P \int_0^{\infty}\frac{d\omega}{\pi}J(\omega)\frac{\omega}{\omega^2-\omega_0^2}
\end{equation}
and
\begin{equation}\label{kas}
\kappa_s=\frac{J(\omega_0)}{2\hbar} \, .
\end{equation}

The time rates of change (\ref{eom2}) include terms of zeroth, first and second order in the system-bath interaction with higher order terms being disregarded. The first-order terms with the random force operator $\xi(t)$ 
can be rewritten. When we disregard the second-order terms, the evolution equations become
\begin{eqnarray}\nonumber\label{eom3}
\hbar   \dot \sigma_x(t)   &=&  [\xi(t)+ F_{\rm eff}(t)]   \sigma_y(t)  \, ,
\\
\hbar \dot \sigma_y(t)   &=&\hbar\omega_0\,  \sigma_z (t)   - [\xi(t)+ F_{\rm eff}(t)]  \sigma_x(t)  \, ,
\\\nonumber
\hbar \dot \sigma_z(t)   &=& -\hbar\omega_0\,   \sigma_y(t)  \, ,
\end{eqnarray}
which are formally solved by
\begin{eqnarray}\label{fsol}
\sigma_x(t)&=&\frac{1}{\hbar}\int_0^tds\,[\xi(s)+ F_{\rm eff}(s)]   \sigma_y(s)\, , \\ \nonumber
\sigma_y(t)&=&\sigma_y(0)\cos(\omega_0t)+\sigma_z(0)\sin(\omega_0t) \\ \nonumber
&&-\frac{1}{\hbar}\int_0^tds\,\cos(\omega_0(t-s))  [\xi(s)+ F_{\rm eff}(s)]  \sigma_x(s)  \, ,\\ \nonumber
\sigma_z(t)&=&\sigma_z(0)\cos(\omega_0t)-\sigma_y(0)\sin(\omega_0t)  \\ \nonumber
&&
+\frac{1}{\hbar}\int_0^tds\,\sin(\omega_0(t-s))   [\xi(s)+ F_{\rm eff}(s)]  \sigma_x(s)\, .
\end{eqnarray}
These spin operators of first order in the system-bath coupling can now be used to transform the terms $\xi(t)\sigma_{x,y}(t)$ in the equations of motion (\ref{eom2}). Since $\xi(t)$ is of first order,  it is consistent, within the second-order approximation, to replace the spin operators multiplied by $\xi(t)$ by their formal first-order solutions (\ref{fsol}). 

Specifically, we shall consider the average spin variables. The equations of motion then contain terms of the form $\langle \xi(t)\sigma_{x,y}(t)\rangle$. Within the second-order approximation, we find
\begin{eqnarray}\label{xix}\nonumber
\langle \xi(t)\sigma_x(t)\rangle&=&\frac{1}{\hbar}\int_0^tds\,\langle \xi(t)[\xi(s)+ F_{\rm eff}(s)]   \sigma_y(s)\rangle\\
&=&\frac{1}{\hbar}\int_0^tds\,C(t-s)\langle  \sigma_y(s)\rangle \, .
\end{eqnarray}
To obtain the latter result, we have disregarded terms of higher than second order in the system-bath coupling as well as terms where the driving force $F_{\rm ext}(t)$ is multiplied by second-order terms, which is appropriate for weak driving. Furthermore, we have inserted the noise correlation function (\ref{xixi}). Likewise, we obtain
\begin{equation}\label{xiy}
\langle \xi(t)\sigma_y(t)\rangle=-\frac{1}{\hbar}\int_0^tds\,\cos(\omega_0(t-s))C(t-s) \langle \sigma_x(s)\rangle \, .
\end{equation}
The expressions in Eqs.\ (\ref{xix}) and (\ref{xiy}) are second-order retardation terms, so that we may again employ the relations (\ref{evo0}) to remove the retardation of the spin variables. This leads in the long time limit to
\begin{eqnarray}\nonumber\label{xisig}
\langle \xi(t)\sigma_x(t)\rangle&=&\hbar\gamma_c \langle  \sigma_y(t)\rangle -\hbar\gamma_s\langle\sigma_z(t)\rangle\, ,\\
\langle \xi(t)\sigma_y(t)\rangle&=&-\hbar\gamma_c\langle\sigma_x(t)\rangle\, ,
\end{eqnarray}
where we have introduced the coefficients
\begin{equation}
\gamma_c=\gamma_c'+i\gamma_c''=\frac{1}{\hbar^2}\int_0^\infty dt\, C(t)\cos(\omega_0 t)
\end{equation}
and
\begin{equation}
\gamma_s=\gamma_s'+i\gamma_s''=\frac{1}{\hbar^2}\int_0^\infty dt\, C(t)\sin(\omega_0 t) \, .
\end{equation}
Using the form (\ref{Coft}) of the noise correlation function, one obtains for the damping coefficients the results (\ref{gacr}) -- (\ref{gasi}) in the main text. In view of Eqs.~(\ref{kac}) and (\ref{kas}), these results imply
\begin{equation}\label{kaga}
\kappa_c=-\gamma'_c\,,\qquad \kappa_s=-\gamma''_s\, .
\end{equation}
 By virtue of the Eqs.\ (\ref{xisig}) and (\ref{kaga}), we obtain from the equations of motion (\ref{eom2}) the evolution equations (\ref{eomav})  for the average spin values.


\end{document}